\documentclass[]{article} 
\usepackage{arxiv}
\usepackage{times}
\usepackage{graphicx}
\usepackage{amssymb}
\usepackage{url,hyperref}
\usepackage[T1]{fontenc}

\usepackage{microtype}
\usepackage{graphicx}
\usepackage{subfigure}
\usepackage{booktabs}

\usepackage{amsmath}
\usepackage{amssymb}
\usepackage{mathtools}
\usepackage{amsthm}
\usepackage{bbm}
\usepackage{bm}
\usepackage{multirow}
\usepackage{makecell}
\usepackage{caption}
\usepackage{algorithm} 
\usepackage{algorithmic}

\title{Retroformer: Pushing the Limits of End-to-end Retrosynthesis Transformer}

\author{Yue Wan, Benben Liao, Chang-Yu Hsieh$^1$, Shengyu Zhang$^2$ \\ Tencent Quantum Laboratory \\
Shenzhen, China \\
\texttt{kimhsieh@tencent.com}$^1$ \\ \texttt{shengyzhang@tencent.com}$^2$
}
\date{}

\begin{document}
\maketitle
\thispagestyle{empty}

\begin{abstract}
Retrosynthesis prediction is one of the fundamental challenges in organic synthesis. The task is to predict the reactants given a core product. With the advancement of machine learning, computer-aided synthesis planning has gained increasing interest. Numerous methods were proposed to solve this problem with different levels of dependency on additional chemical knowledge. In this paper, we propose Retroformer, a novel Transformer-based architecture for retrosynthesis prediction without relying on any cheminformatics tools for molecule editing. Via the proposed local attention head, the model can jointly encode the molecular sequence and graph, and efficiently exchange information between the local reactive region and the global reaction context. Retroformer reaches the new state-of-the-art accuracy for the end-to-end template-free retrosynthesis, and improves over many strong baselines on better molecule and reaction validity. In addition, its generative procedure is highly interpretable and controllable. Overall, Retroformer pushes the limits of the reaction reasoning ability of deep generative models. 
\end{abstract}

\section{Introduction}
\label{intro}

Retrosynthesis \cite{retro_first} is one of the major building blocks in organic synthesis, which aims to discover valid and efficient synthetic routes (i.e., reactants) given a target molecule (i.e., product). It is crucial for the pharmaceutical industry as one of the main challenges for drug discovery is to efficiently synthesize novel and complex compounds in the laboratory \cite{organic_synthesis}. 

\begin{figure}[]
    \centering
    % trim={<left> <lower> <right> <upper>}
    \includegraphics[width=0.5\columnwidth, trim={5.2cm 8.5cm 5cm 4cm},clip]{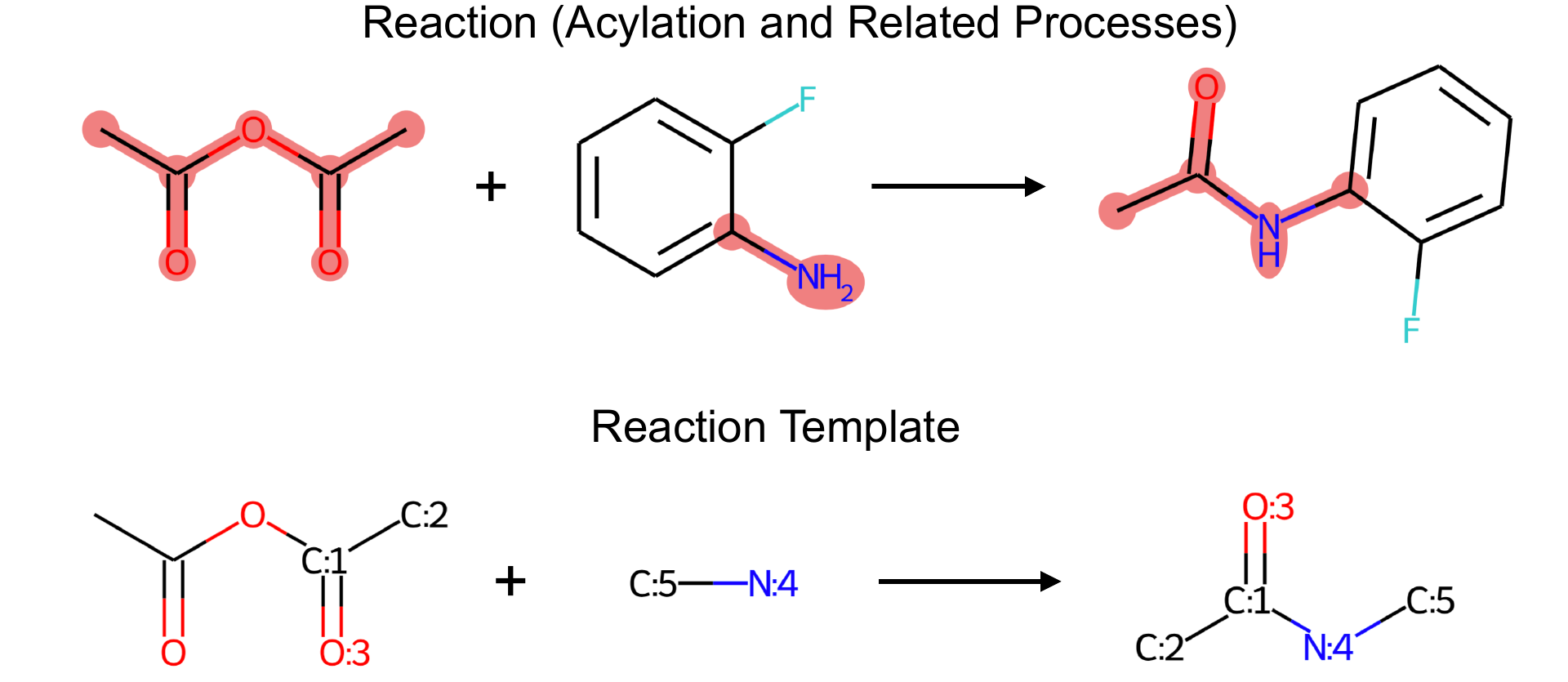}
    \caption{Sample reaction (top) and its corresponding reaction template (bottom).}
    \label{fig:rxn_tpl_sample}
\end{figure}

Recently, computer-aided synthesis planning has gained vast attention for its potential to save a tremendous amount of time and efforts from traditional retrosynthesis approaches. Various machine learning approaches were proposed with different levels of dependency on additional chemical knowledge. These methods can be categorized into three groups. First, template-based methods \cite{retrosim, gln, localretro} view the retrosynthesis prediction as the template retrieval problem, where a template encodes the core reactive rule (Figure~\ref{fig:rxn_tpl_sample}). After the templates are retrieved, these methods use cheminformatics tools like RDKit \cite{rdkit} to build up full reactions from the templates. Despite the state-of-the-art accuracy and guaranteed molecule validity, these methods are limited to the scope of the existing template database. In contrast,  template-free methods, the second class, use deep generative models to directly generate the reactants given the product. Since molecule can be represented by both the graph and the SMILES sequence, existing approaches reframe the retrosynthesis into either sequence-to-sequence \cite{templatefreemulti, retro_diverse_pretrain, SCROP, retro_at, GTA, tied_transformer} or graph-to-sequence problem \cite{graph2smiles}. These generative methods do not rely on any additional chemical knowledge and can perform chemical reasoning within a larger reaction space. The third class is semi-template-based methods, which combine the advantages of both the generative models and the additional chemical knowledge. Conventional frameworks \cite{retroxpert, graph2graph, graphretro, retroprime} in this category follow the same idea: They first identify the reactive bond and convert the product into synthons by RDKit. Then, another model completes synthons into reactants. These methods are competitive in accuracy and are interpretable by their stage-wise nature.

In this work, we are interested in the template-free generative approach for retrosynthesis prediction. Existing methods fail to fully explore the potential of deep generative model in terms of reaction reasoning, and we argue that the end-to-end Transformer-based \cite{transformer} architecture can reach the same competitive benchmark accuracy as well as good validity and interpretability. We propose Retroformer, a novel end-to-end retrosynthesis Transformer that introduces a special attention head. It is able to jointly encode the sequential and graphical information of the molecule and allow efficient information exchange between the local reactive region and the global reaction context. The generative process is also sensitive to the exact reactive region. Our end-to-end model does not rely on any additional helps from the cheminformatic tools for molecule editing. Experiments show that our model can improve over the vanilla Transformer by 12.5\% and 14.4\% top-10 accuracy in the reaction class known and unknown settings, respectively. It reaches the new state-of-the-art accuracy for template-free methods and is competitive against both template-based and semi-template-based methods. It also enjoys better molecule and reaction validity compared to strong baseline models. The model is highly interpretable and controllable for downstream usage. Our contributions are summarized as:
\begin{itemize}
    \item We propose Retroformer, a novel Transformer-based architecture that introduces the local attention head, to push the limits of the reaction reasoning ability of deep generative models in retrosynthesis prediction.
    \item The proposed method reaches 64\% and 53.2\% top-1 accuracy for reaction class known and unknown settings, respectively, which is the new state-of-the-art performance for template-free retrosynthesis.
    \item The proposed method further improves the top-10 molecule and reaction validity by 23.6\% and 22.0\%, respectively, compared to the vanilla retrosynthesis Transformer.
\end{itemize}

\section{Related Work}
\subsection{Retrosynthesis Prediction} 
Existing methods in retrosynthesis prediction can be grouped into three categories: template-based, template-free, and semi-template-based. The reaction template encodes the core reactive rules. As shown in Figure~\ref{fig:rxn_tpl_sample}, a conventional template tells the potential reactive region within the molecule, as well as its potential chemical transformation. These templates are either expert-defined or automatically extracted by algorithms. In this work, we strictly differentiate the three categories by the levels of dependency on additional chemical knowledge during inference. 

\textbf{Template-based} methods rely on an external template database. Since the template is a more efficient and interpretable representation for reactions \cite{templatecorr, neuraltpl}, a large body of works \cite{retrosim, gln, localretro} focus on capturing the reactive scores between the molecules and templates. Retrosim \cite{retrosim} uses molecule fingerprint similarity to rank the candidate templates. GLN \cite{gln} and LocalRetro \cite{localretro} use graph neural network (GNN) to capture the molecule-template and atom/bond-template relationship, respectively. Despite their state-of-the-art top-$k$ accuracy, all template-based methods suffer from the incomplete coverage issue and do not scale well. 

\textbf{Template-free} methods, in contrast, adopt deep generative models to directly generate the reactants molecules. Besides graph, molecules can be represented using SMILES sequence. Existing works \cite{templatefreemulti, retro_diverse_pretrain, SCROP, retro_at, GTA, tied_transformer} take advantage of the Transformer \cite{transformer} architecture and reframe the problem as the sequence-to-sequence translation from product to reactants. Graph2SMILES \cite{graph2smiles} replaces the original sequence encoder with a graph encoder to ensure the permutation invariance of SMILES. These methods rely on little additional chemical knowledge for inference. However, chemical validity can be a huge concern because validity is often not part of the training objective. Another factor is the ignorance of graphical structure during the sequence generation. Also, generated outcomes from beam search often suffer from the diversity issue \cite{diverse_bs}, which is another practical concern for retrosynthesis. 

\begin{figure*}[!ht]
    \centering
    % trim={<left> <lower> <right> <upper>}
    \includegraphics[width=0.9\textwidth, trim={2.5cm 7cm 1cm 4cm},clip]{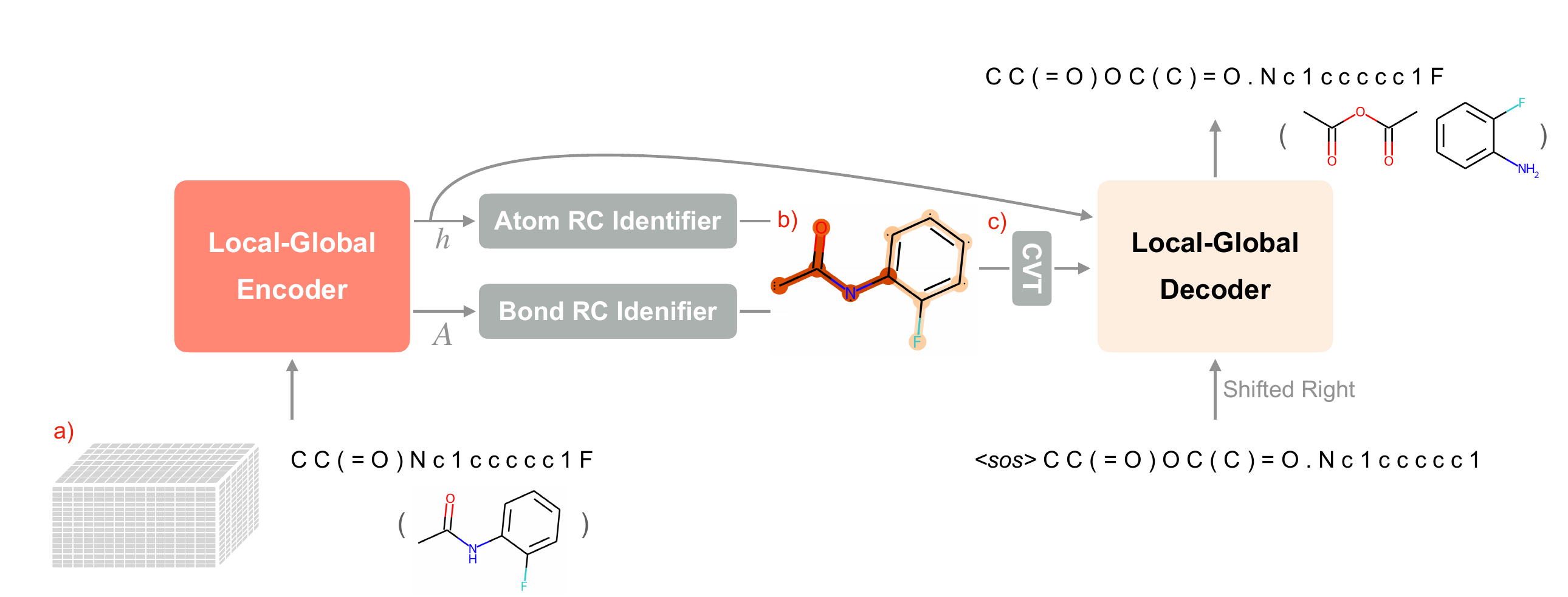}
    \caption{Architecture overview. The model takes molecular SMILES $S$ and a) bond feature matrix $A$ as inputs. Besides the encoder outputs $h$, the b) predicted reaction center $S_{rc}$ is c) converted to attention masking and passed to the decoder.}
    \label{fig:overview}
\end{figure*}

\begin{figure}[!ht]
    \centering
    % trim={<left> <lower> <right> <upper>}
    \includegraphics[width=0.65\columnwidth, trim={2cm 6.3cm 9.5cm 4.5cm},clip]{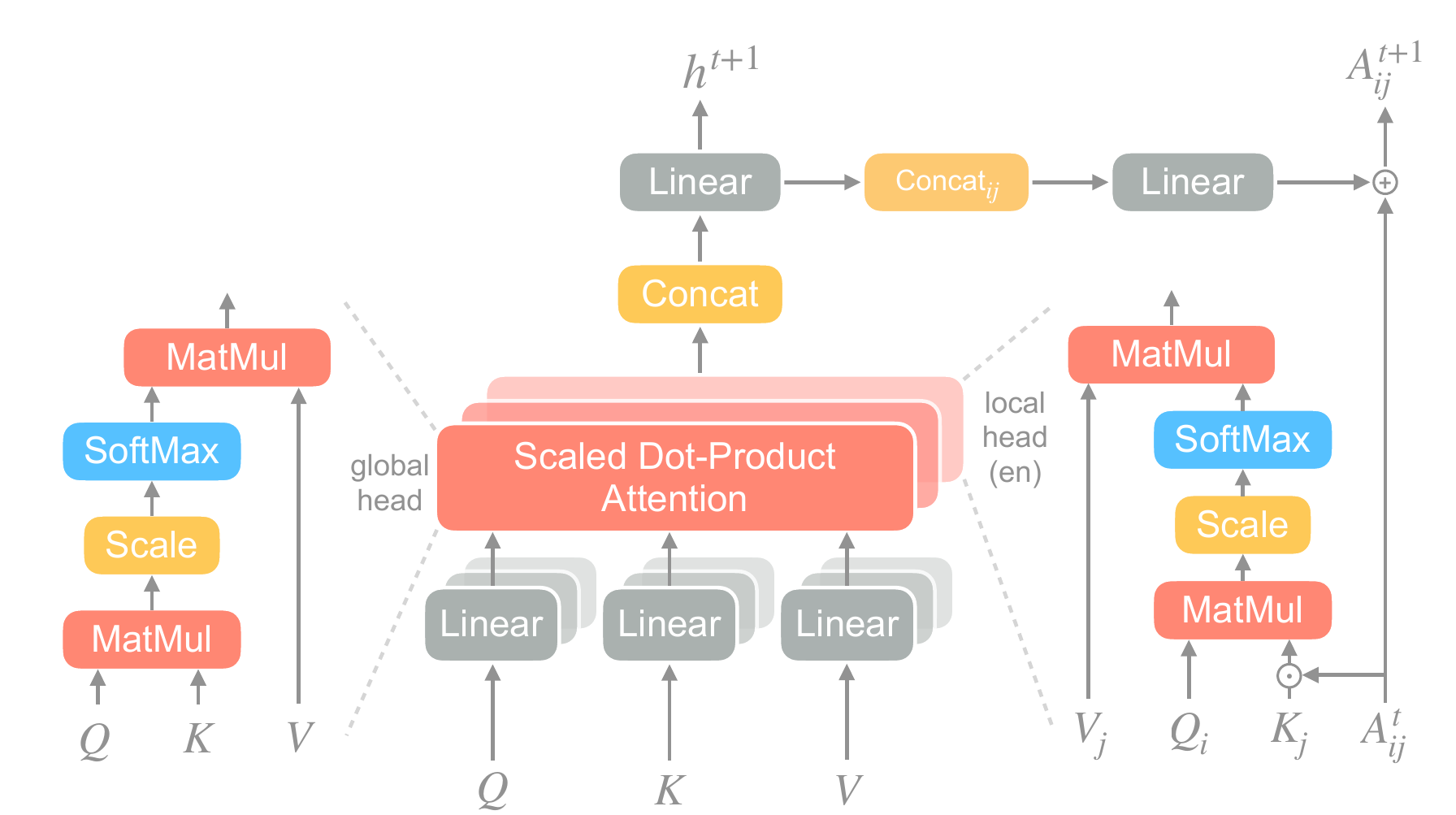}
    \caption{Local-global self attention head in encoder.}
    \label{fig:attention}
\end{figure}

\textbf{Semi-template-based} methods combine the advantage of both the generative models and additional chemical knowledge. In this work, we strictly categorize generative models which require additional help from RDKit \cite{rdkit} for molecule editing into this method group. Most existing works \cite{retroxpert, graph2graph, graphretro, retroprime} approach the task by a two-stage procedure. Despite their architecture differences between GNN and Transformer, they follow the same idea: They first convert the product into synthons by breaking the reactive bond via RDKit, then complete the synthons into reactants by either leaving groups selection \cite{graphretro}, graph generation \cite{graph2graph}, or SMILES generation \cite{retroxpert, retroprime}. In contrast, MEGAN \cite{MEGAN} reframes the generative procedure as a sequence of graph edits that are completed by RDKit. 

Besides different architecture designs, self-supervised molecule pretraining is also shown to be effective in retrosynthesis prediction. DMP-fusion \cite{dualview} pretrains the molecule with a dual view of SMILES and graph. Chemformer \cite{chemformer} applies masked SMILES modeling to learn the molecule representation.

\subsection{Graph Transformer} 
The introduction of Transformer into the graph domain has gained increasing interest. The global receptive fields of the self-attention and the local message passing of the graph neural network are inherently complementary and compatible. Attempts have been made in ways of incorporating graph information into the self-attention computation \cite{graphormer, MAT, R-MAT} and integrating the conventional graph neural networks \cite{nmpn} with Transformer architecture \cite{generalizationtograph, compt}.

\section{Preliminary}
Let $S = [s_1, s_2, ... s_n]$ be the molecular SMILES sequence with $n$ number of tokens. Let $G_{mol}=(V_m, E_m)$ be the molecular graph. It is formed by $|V_m|$ number of atoms with $|E_m|$ number of bonds. For computation convenience, we further introduce the SMILES graph $G_{smi}=(V_s, E_s)$. $V_s$ is made up of all the SMILES tokens, including the atom tokens (e.g., ``C'', ``O'') as well as the other special tokens (e.g., ``='', ``1''): $V_m \subseteq V_s = S$. In $G_{smi}$, the special tokens are treated as trivial nodes with no neighbors. Its edge $E_s$ represents the graphical connections between atom tokens, which is essentially the same as bond connections $E_m$ in $G_{mol}$. In general, $G_{smi}$ is a larger but sparser graph compared to $G_{mol}$. The introduction of $G_{smi}$ is merely to ensure the alignment relationship between the atoms in graph and the tokens in SMILES. 

\section{Retroformer}
We propose Retroformer, a novel Transformer-based model that is able to perform interpretable retrosynthesis prediction in an end-to-end manner. We propose a special type of local attention head that can support efficient information exchange between the local region of reactive importance and the global reaction context. Its generative procedure is also sensitive to the exact local region. The overall training and inference can be done in an end-to-end manner. It is a fully template-free method without any additional dependency on RDKit for molecule editing. The overall architecture contains an encoder, a decoder, and two reaction center identifiers. We also propose to use SMILES alignment and on-the-fly data augmentation as two additional training strategies.

\subsection{Local-Global Encoder with Edge Update}
Since molecular graph can provide additional information on top of the SMILES sequence, our encoder takes both $S$ sequence and $G_{smi}$ (i.e., adjacency matrix and bond feature) as inputs. The bond features we considered are listed in Appendix \ref{sec:bond_feature}. Different from the existing graph Transformers \cite{graphormer, MAT, R-MAT} that compute graph self-attention within the entire module, our model encodes the graph information at the head level. We specify two types of attention heads: global head and local head. The global head is the same as the vanilla self-attention head, where its receptive field is the entire SMILES sequence. The local head, on the other hand, considers the topological structure of the molecule. The receptive field of the individual token is restricted to its one-hop neighborhood, which is similar to \cite{mg-bert}. In addition, we perform element-wise multiplication between the key vector and the edge feature to incorporate the bond information into the calculation. The roll-out form of the local head self-attention at layer $l$ for the $i^{th}$ token is formulated as: 
\begin{align}
    % {x_{i}^{l+1}}_{global} = \sum_{j \in S} \sigma(\frac{\bm{q_i} \bm{k_j}^T}{\sqrt{d}}) \bm{v_j} \\
    {x_{i}^{l+1}}_{local} & = \sum_{j \in N(i)} \sigma(\frac{{q_i} ({A_{ij}^l} \odot {k_j})^T}{\sqrt{d}}) {v_j} \\
    [{q_i}, {k_j}, {v_j}] & = [{h_i^l} W^Q, {h_j^l} W^K, {h_j^l} W^V] \notag
\end{align}

where $A$ is the bond feature matrix, $W^Q, W^K, W^V$ are the projection matrix for query $q$, key $k$, and value $v$, and $\sigma$ is the softmax operation. The computed representations from the global and local heads are then concatenated along the hidden dimension and passed to a linear layer, which represents the updated token features $h^{l+1}$. Meanwhile, the edge update module is a fully connected layer (FFN) that takes the concatenation of the updated features of the receiving and sending tokens as inputs:
\begin{align}
    h^{l+1} & = \text{Linear}([{x^{l+1}}_{global} ; {x^{l+1}}_{local}]) \\
    A_{ij}^{l+1} & = A_{ij}^l + \text{FFN}([h^{l+1}_i; h^{l+1}_j])
\end{align}

The integration of the local, global attention heads, and the edge update module allows the model to efficiently exchange information between the local region and global molecular context. Same as the vanilla Transformer \cite{transformer}, layer normalization and residual connection are enforced between encoder layers. The final encoder outputs are the updated token representation $h$ and the bond representation $A$. 

\subsection{Reaction Center Detection}
A reaction center represents the group of atoms and bonds that are contributing factors to the chemical transformation. However, existing semi-template-based methods \cite{retroxpert, graphretro, graph2graph, retroprime} simplify this concept as the reactive bond. We argue that this simplification will lead to information loss of the reaction context. These methods also cannot perform retrosynthesis in an end-to-end manner, since they rely on RDKit to convert the product into synthons. Instead, Retroformer predicts the reactive probability $P_{rc}(.)$ of each atom and bond and infers the reactive region of $S$ as the attention receptive field for the decoder. In other words, the detected reaction center $S_{rc}$ is a subset of $S$. 

Figure~\ref{fig:overview}b shows a heat map visualization of the predicted reactive probability. It is done by two fully connected layers named Atom RC Identifier and Bond RC Identifier: 
\begin{align}
    P_{rc}(s_i) & = \sigma \big(\text{FFN}_{\text{atom}}(h_i)\big), s_i \in V_m \\
    P_{rc}(e_{ij}) & = \sigma \big(\text{FFN}_{\text{bond}}(A_{ij})\big), e_{ij} \in E_m
\end{align}

We will show in Section \ref{sec:qualitative} that the learned reaction center can be easily visualized and matched with chemical heuristics. We then convert the atom and the bond reactive probability into the reactive indicator of tokens in $S_{rc}$ by either one of the following two strategies:

\begin{itemize}
    \item \textit{naive}: we naively set a token as reactive if it exists in a reactive edge (i.e., $P_{rc}(e) > 0.5$) and is reactive itself (i.e., $P_{rc}(s) > 0.5$). Note that the special tokens are guaranteed to be non-reactive. This strategy is used at both training and inference stages. 
    \item \textit{search}: we conduct a subgraph search on the molecular graph and rank the subgraphs by their reaction center score: $\sum_{s_i \in S_{rc}} \log P_{rc}(s_i) + \sum_{s_i,s_j \in S_{rc}} \log P_{rc}(e_{ij})$. Only atoms with $P_{rc}(s) > \alpha_{atom}$ and bonds with $P_{rc}(e) > \alpha_{bond}$ are considered in the search to reduce the computational time. Detailed algorithm is described in Appendix \ref{sec:reaction_center_search}. Then, top-$n$ subgraphs are selected as reaction center candidates. The model then generates $k/n$ reactants for each reaction center, where $k$ is the total number of predicted reactants. The final results are ranked by the sum of the reaction center score and the generative score. This strategy is only used at inference stage.
    
\end{itemize}

\begin{figure}[t]
    \centering
    % trim={<left> <lower> <right> <upper>}
    \includegraphics[width=0.7\columnwidth, trim={1.5cm 12.3cm 2cm 1cm},clip]{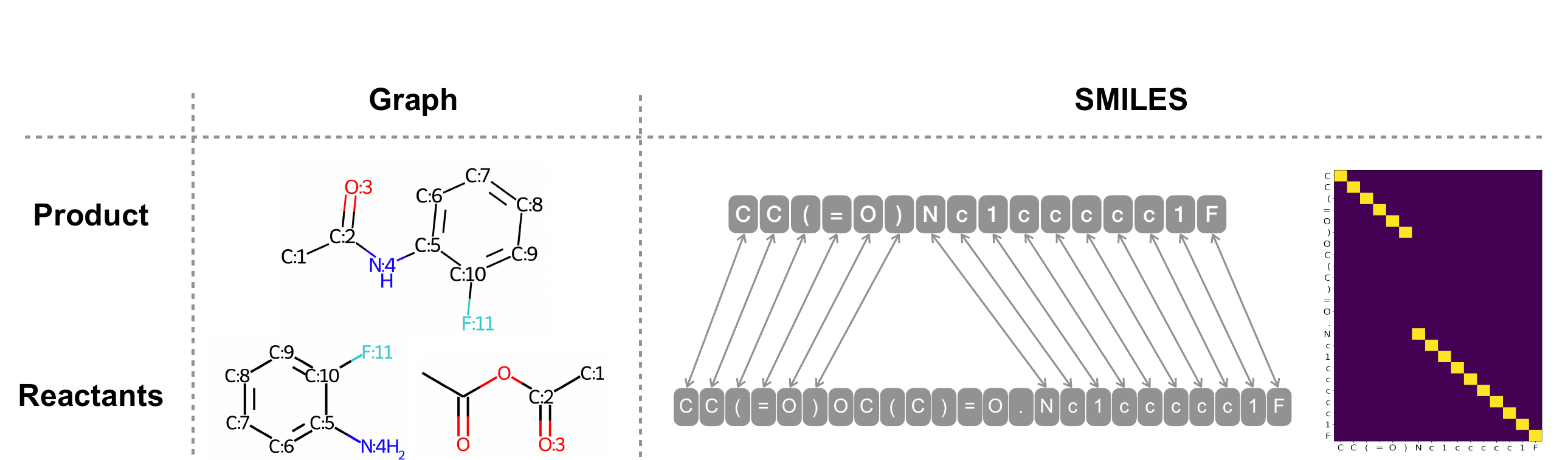}
    \caption{Token and atom alignment between product and reactants in SMILES and graph representations, respectively. The right most figure is the ground truth alignment matrix.}
    \label{fig:alignment}
\end{figure}

\subsection{Local-Global Decoder}
The decoder takes its generative outcomes from the previous step, the encoder outputs $h$, and the reaction center $S_{rc}$ as inputs. Similar to the encoder, we also introduce two different heads in its cross-attention module. The global head is the same as the vanilla head. The local head, on the contrary, is only visible to the detected reaction center $S_{rc}$. It computes the sparse cross-attention instead of the full cross-attention. 
\begin{align}
    {y_{i}^{l+1}}_{local} & = \sum_{s_j \in S_{rc}} \sigma(\frac{{q_i} {k_j}^T}{\sqrt{d}}) {v_j} \\
    [{q_i}, {k_j}, {v_j}] & = [{g_i^l} W^Q, {h_j} W^K, {h_j} W^V] \notag
\end{align}

Same as the encoder, the computed representations from the global and local heads are then concatenated along the hidden dimension and passed to a linear layer, which represents the representation $g^{l+1}$. It essentially converts the decoder into a conditional generative module.
\begin{align}
    g^{l+1} & = \text{Linear}([{y^{l+1}}_{global} ; {y^{l+1}}_{local}])
\end{align}

\subsection{SMILES Alignment}
SMILES alignment is an additional learning task of Retroformer. Similar to machine translation, the SMILES sequences of the source and the target molecules are often partially aligned. A large portion of the molecules remains unchanged during the reaction. Figure~\ref{fig:alignment} shows this alignment relationship in both graph and SMILES representations. The node alignment between graphs (i.e., atom mapping) can be easily converted into token alignment between SMILES. Detailed substring matching algorithm with atom-mapping is described in Appendix \ref{sec:get_alignment}.

Inspired by the effectiveness of the guided attention in \cite{align_transformer}, we introduce the attention guidance loss between the ground truth alignment and the attention weights from the decoder's global cross-attention heads. We treat the computed cross-attention at each decoder step as a probability distribution and impose a label smoothing loss \cite{label_smoothing}. It is a soft cross entropy loss with the label smoothing technique and is shown to be effective in classification performance. Hypothetically speaking, this guided attention can encourage the model to understand chemical reactions more efficiently.

\subsection{Data Augmentation}
We follow the same data augmentation tricks used in \cite{GTA, retro_at} for the SMILES generative models, which are the SMILES permutation of the product and the order permutation of reactants. However, instead of expanding the training dataset off-the-shelf, we choose to perform the augmentation on-the-fly. At each iteration, there is a probability of 50\% to permute the product SMILES and another probability of 50\% to permute the reactants order. This dynamic permutation allows the model to focus more on the canonical SMILES and use the permuted SMILES for regularization. 

\subsection{Loss}
The training schema of Retroformer can be viewed as an end-to-end multi-task learning. The overall loss is made up of four parts: $\mathcal{L} = \mathcal{L}_{LM} + \mathcal{L}_{RC_{bond}} + \mathcal{L}_{RC_{atom}} + \mathcal{L}_{AG}$, where $\mathcal{L}_{LM}$ is the language modeling objective, $\mathcal{L}_{RC_{*}}$ is the reactive probability loss, and $\mathcal{L}_{AG}$ is the SMILES attention guidance loss.  

\begin{table*}[t]
\centering
\caption{Top-$k$ accuracy for retrosynthesis prediction on USPTO-50K. * indicates the model with SMILES augmentation. For comparison purpose, the Aug. Transformer is evaluated without the test augmentation. Best performance is in \textbf{bold}.}
\begin{tabular}{lccccccccc}
\Xhline{1pt}
\multicolumn{1}{c}{\multirow{3}{*}{\textbf{Model}}} & \multicolumn{9}{c}{\textbf{Top-k accuracy (\%)}}                                                                                                                                                 \\ \cline{2-10} 
\multicolumn{1}{c}{}                                & \multicolumn{4}{c}{\textbf{Reaction class known}}                                                         & \textbf{}            & \multicolumn{4}{c}{\textbf{Reaction class unknown}}           \\ \cline{2-5} \cline{7-10} 
\multicolumn{1}{c}{}                                & 1                        & 3                        & 5                        & 10                       &                      & 1             & 3             & 5             & 10            \\ \hline
\textbf{Template-Based}                             &                          &                          &                          &                          &                      &               &               &               &               \\ \hline
GLN  \cite{gln}                                               & \textbf{64.2}            & 79.1                     & 85.2                     & 90.0                     &                      & 52.5          & 69.0          & 75.6          & 83.7          \\
LocalRetro  \cite{localretro}                                        & 63.9                     & \textbf{86.8}            & \textbf{92.4}            & \textbf{96.3}            &                      & \textbf{53.4} & \textbf{77.5} & \textbf{85.9} & \textbf{92.4} \\ \hline
\textbf{Template-Free}                              &                          &                          &                          &                          &                      &               &               &               &               \\ \hline
Transformer                                         & 57.1                     & 71.5                     & 75.0                     & 77.7                     &                      & 42.4          & 58.6          & 63.8          & 67.7          \\
SCROP \cite{SCROP}                                              & 59.0                     & 74.8                     & 78.1                     & 81.1                     &                      & 43.7          & 60.0          & 65.2          & 68.7          \\
Tied Transformer \cite{tied_transformer}                                   & -                        & -                        & -                        & -                        &                      & 47.1          & 67.1          & 73.1          & 76.3          \\
Aug. Transformer* \cite{retro_at}                                   & -                        & -                        & -                        & -                        &                      & 48.3          & -             & 73.4          & 77.4          \\
GTA* \cite{GTA}                                               & -                        & -                        & -                        & -                        &                      & 51.1          & 67.6          & 74.8          & 81.6          \\
Graph2SMILES \cite{graph2smiles}                                       & -                        & -                        & -                        & -                        &                      & {52.9}          & 66.5          & 70.0          & 72.9          \\
Retroformer$_{\text{base}}$ (Ours)                                               & \multicolumn{1}{l}{61.5} & \multicolumn{1}{l}{78.3} & \multicolumn{1}{l}{82.0} & \multicolumn{1}{l}{84.9} & \multicolumn{1}{l}{} & 47.9          & 62.9          & 66.6          & 70.7  \\
Retroformer$_{\text{aug}}$* (Ours)                                         & \textbf{64.0}            & 81.8                     & 85.4                     & 88.3                     &                      & {52.9}          & 68.2          & 72.5          & 76.4          \\
Retroformer$_{\text{aug}}+$* (Ours)                                                & \textbf{64.0}            & \textbf{82.5}            & \textbf{86.7}            & \textbf{90.2}            &                      & \textbf{53.2}          & \textbf{71.1} & \textbf{76.6} & \textbf{82.1} \\ \hline
\textbf{Semi-Template-Based}                        &                          &                          &                          &                          &                      &               &               &               &               \\ \hline
RetroXpert* \cite{retroxpert}                                        & 62.1                     & 75.8                     & 78.5                     & 80.9                     &                      & 50.4          & 61.1          & 62.3          & 63.4          \\
G2G \cite{graph2graph}                                                & 61.0                     & 81.3                     & {86.0}            & {88.7}            &                      & 48.9          & 67.6          & 72.5          & 75.5          \\
GraphRetro  \cite{graphretro}                                         & 63.9                     & 81.5                     & 85.2                     & 88.1                     &                      & \textbf{53.7} & 68.3          & 72.2          & 75.5          \\
RetroPrime* \cite{retroprime}                                        & \textbf{64.8}            & {81.6}            & 85.0                     & 86.9                     &                      & 51.4          & \textbf{70.8} & {74.0} & {76.1} \\ 
MEGAN \cite{MEGAN}
& 60.7 & \textbf{82.0} & \textbf{87.5} & \textbf{91.6} &  & 48.1 & 70.7  & \textbf{78.4}  & \textbf{86.1} \\ \Xhline{1pt}
\end{tabular}
\end{table*}

\section{Experiments}
\paragraph{Data}
We use the conventional retrosynthesis benchmark dataset USPTO-50K \cite{uspto50k} to evaluate our method. It contains 50016 atom-mapped reactions that are grouped into 10 reaction classes. We use the same data split as \cite{retrosim}. We canonicalize the molecule SMILES with atom mapping following the same protocol given in \cite{graphretro}. We then use RDChiral \cite{rdchiral} to extract the ground truth reaction center and use the algorithm in Appendix \ref{sec:get_alignment} to extract the ground truth SMILES token alignment. 

\paragraph{Evaluation}
We adopt the conventional top-$k$ accuracy of the full reactants to evaluate the retrosynthesis performance. We also evaluate the top-$k$ validity of the generated routes. For molecule validity, we treat a candidate as valid if RDKit \cite{rdkit} can successfully identify the molecule SMILES. The top-$k$ validity is calculated as: $Valid(k) = \frac{1}{N \times k} \sum_1^N \sum_1^k \mathbbm{1}({\text{SMILES is valid}})$. We further evaluate our method with the round-trip accuracy \cite{metrics}, which is an approximation metric for reaction validity. It measures the percentage of predicted reactants that can lead back to the original product. We take the pretrained Molecule Transformer \cite{molecule_transformer} as the oracle reaction prediction model because of its state-of-the-art performance. Our top-$k$ round-trip accuracy calculation is slightly different from the definition adopted by RetroPrime \cite{retroprime} and LocalRetro \cite{localretro}: $RoundTrip(k) = \frac{1}{N \times k} \sum_1^N \sum_1^k \mathbbm{1} (\text{Reach Ground Truth Product})$. 

\paragraph{Baseline} We take GLN \cite{gln} and LocalRetro \cite{localretro} as two strong baseline models from the \textit{template-based} group. We take SCROP \cite{SCROP}, Tied Transformer \cite{tied_transformer}, Augmented Transformer \cite{retro_at}, GTA \cite{GTA}, and Graph2Smiles \cite{graph2smiles} as the baseline models from the \textit{template-free} group. We also train a vanilla retrosynthesis Transformer from scratch using OpenNMT \cite{opennmt}. We take RetroXpert \cite{retroxpert}, G2G \cite{graph2graph}, GraphRetro \cite{graphretro}, RetroPrime \cite{retroprime}, and MEGAN \cite{MEGAN} as strong \textit{semi-template-based} baselines. We do not include the pretraining approach in the performance comparison. We experiment with three variants of the proposed model: Retroformer$_{\text{base}}$ represents the model with no data augmentation and the \textit{naive} reaction center detection strategy; Retroformer$_{\text{aug}}$ represents the model with data augmentation and the \textit{naive} strategy; Retroformer$_{\text{aug}}+$ represents the model with data augmentation and the \textit{search} strategy. 

\paragraph{Implementation Details} Built on top of the vanilla Transformer \cite{transformer}, our model has 8 encoder layers and 8 decoder layers. The model is trained using the Adam optimizer \cite{adam} with a fixed learning rate of $1e-4$, and a dropout rate of $0.3$. The embedding dimension is set to 256, and the total amount of heads is set to 8. We split the heads by half for global and local heads. The Retroformer$_\text{base}$ is trained on 1 NVIDIA Tesla V100 GPU for 24 hours.  

\begin{table}[]
\centering
\caption{Top-$k$ SMILES validity for retrosynthesis prediction on USPTO-50K with reaction class unknown. }
\vspace{5pt}
\begin{tabular}{lllll}
\Xhline{1pt}
\multicolumn{1}{c}{\textbf{Model}} & \multicolumn{4}{c}{\textbf{Top-k validity (\%)}}                                                                               \\ \cline{2-5} 
\multicolumn{1}{c}{}               & \multicolumn{1}{c}{1}                & \multicolumn{1}{c}{3}       & \multicolumn{1}{c}{5}       & \multicolumn{1}{c}{10}      \\ \hline
Transformer                        & {97.2}          & {87.9} & {82.4} & {73.1} \\
Graph2SMILES  & \textbf{99.4} & 90.9 & 84.9 & 74.9 \\
RetroPrime & 98.9 & 98.2 & 97.1 & 92.5 \\
Retroformer$_{\text{aug}}$  & {{99.3}}          & {98.5} & {97.2} & {92.6} \\
Retroformer$_{\text{aug}} +$                                & {99.2} & {\textbf{98.5}} & {\textbf{97.4}} & {\textbf{96.7}} \\ \Xhline{1pt}
\end{tabular}
\label{tab:validity}
\end{table}

\begin{table}[]
\centering
\caption{Top-$k$ round-trip accuracy for retrosynthesis prediction on USPTO-50K with reaction class unknown.}
\vspace{5pt}
\begin{tabular}{lcccc}
\Xhline{1pt}
\multicolumn{1}{c}{\textbf{Model}} & \multicolumn{4}{c}{\textbf{Top-k round-trip acc. (\%)}} \\ \cline{2-5} 
\multicolumn{1}{c}{} & 1 & 3 & 5 & 10 \\ \hline
Transformer & 71.9 & 54.7 & 46.2 & 35.6 \\
Graph2SMILES & 76.7 & 56.0 & 46.4 & 34.9 \\
RetroPrime & \textbf{79.6} & 59.6 & 50.3 & 40.4 \\
Retroformer$_{\text{aug}}$ & {{78.6}} & {71.8} & \textbf{67.1} & \textbf{57.6} \\
Retroformer$_{\text{aug}} +$ & {78.9} & \textbf{72.0} & \textbf{67.1} & {57.2} \\ \Xhline{1pt}
\end{tabular}
\label{tab:round_trip}
\end{table}

\begin{table*}[t]
\centering
\caption{Effects of different modules on retrosynthesis performance in reaction class unknown setting.}
\begin{tabular}{cccccclllll}
\Xhline{1pt}
\multicolumn{1}{l}{\textbf{Settings}} &  \multicolumn{5}{c}{\textbf{Modules}}  &  & \multicolumn{4}{c}{\textbf{Top-k accuracy (\%)}} \\ \hline
\multicolumn{1}{l}{} &  \multicolumn{1}{l}{\begin{tabular}[c]{@{}c@{}}Guided$_{\text{last}}$\end{tabular}} & \multicolumn{1}{l}{\begin{tabular}[c]{@{}c@{}}Guided$_{\text{all}}$\end{tabular}} & \multicolumn{1}{l}{\begin{tabular}[c]{@{}c@{}}Local-global \\ Encoder\end{tabular}} & \multicolumn{1}{l}{\begin{tabular}[c]{@{}c@{}}Local-global \\ Decoder\end{tabular}} & \multicolumn{1}{l}{\begin{tabular}[c]{@{}c@{}}Reaction \\ Center Search\end{tabular}} &  & 1 & 3 & 5 & 10 \\ \cline{1-6} \cline{8-11}
$(a)$ & & & $\surd$ & $\surd$ & & & 45.5 & 60.7 & 65.4 & 69.9 \\
$(b)$ & & $\surd$ & $\surd$ & $\surd$ & & & 47.0 & 63.1 & 66.9 & 71.1 \\
$(c)$ & $\surd$ & & $\surd$ & $\surd$ & & & {47.9} & 62.9 & 66.6 & 70.7 \\
$(d)$ & $\surd$ & & & $\surd$ & & & 44.1 & 60.1 & 64.7 & 70.2 \\
$(e)$ & $\surd$ & & $\surd$ & & & & 46.7 & 63.7 & 68.4 & 73.9 \\
$(f)$ & $\surd$ & & $\surd$ & $\surd$ & $\surd$ & & \textbf{48.4} & \textbf{66.8} & \textbf{73.2} & \textbf{78.8}    \\ \Xhline{1pt}
\end{tabular}
\label{tab:ablation}
\end{table*}

\subsection{Performance}
\paragraph{Top-k Accuracy}
With the reaction class known, our augmented model can achieve a 64.0\% top-$1$ and 88.3\% top-$10$ accuracy. It reaches the state-of-the-art performance for template-free methods and is competitive against template-based and semi-template-based methods. It improves over the vanilla retrosynthesis Transformer by 6.9\% top-$1$ and 11.9\% top-$10$, respectively. With the reaction class unknown, our augmented model can achieve a 52.9\% top-$1$ and 76.4\% top-$10$ accuracy. The top-$1$ accuracy reaches the state-of-the-art performance as Graph2SMILES. In addition, Retroformer$_\text{base}$ surpasses the vanilla retrosynthesis Transformer by a large margin in both settings. It demonstrates the promising potential for the deep generative model to perform end-to-end retrosynthesis prediction and reaction space exploration.

We further demonstrate the strength of the reaction center detection. With the top-$n$ subgraphs proposed, Retroformer$_{\text{aug}} +$ can further boost the performance to the new state-of-the-art accuracy for template-free retrosynthesis in both experiment settings. We provide further interpretation of the $search$ strategy in Section~\ref{sec:qualitative} and Appendix~\ref{sec:reaction_center_search}.

\paragraph{Top-k SMILES Validity}
We take the vanilla retrosynthesis Transformer, Graph2SMILES, and RetroPrime as strong SMILES generative baselines for validity comparison with our model. We do not include template-based methods in this evaluation since molecule SMILES built from templates are guaranteed to be valid. As we mentioned before, SMILES generative models are more likely to struggle with the validity issue. Without knowing the proper reaction center, the models may modify the molecule fragments that are distant from the core reactive region, which is chemically trivial. As shown in Table~\ref{tab:validity}, both of our model variants enjoy better molecule validity than others. It improves the top-10 validity over the vanilla Transformer by 23.6\%. It shows that being aware of the local reactive region can encourage the model to avoid errors that propagate via the non-reactive regions.

\paragraph{Top-k Round-trip Accuracy}
To measure the reaction validity, we take the pretrained Molecule Transformer \cite{molecule_transformer} as the oracle reaction prediction model to measure the percentage of top-$k$ proposed synthetic routes that can lead back to the ground truth product. Table~\ref{tab:round_trip} shows the performance comparison of the round-trip accuracy. It shows that our method improves over the existing methods by a large margin. Our model exceeds the vanilla Transformer by 22.0\% top-10 round-trip accuracy, and it also improves over RetroPrime by 12.2\%. It shows that our model is more likely to propose valid and efficient synthetic routes for downstream usage. 

\subsection{Ablation Study}
We further conduct ablation study to evaluate the effects of each component on retrosynthesis performance. As for the guided alignment loss, we experiment with two settings: Guided$_{\text{all}}$: the alignment loss is enforced at the first global heads of all decoder layers; Guided$_{\text{last}}$: the loss is enforced at the first global head of the last decoder layer. 

Table~\ref{tab:ablation} shows that all proposed components are necessary for Retroformer$_\text{base}$ to reach the best retrosynthesis performance. The improvement from $(a)$ to $(b)$ and $(a)$ to $(c)$ shows that the model can better capture the reaction knowledge from learning the SMILES alignment. We choose $(c)$ over $(b)$ as our final alignment loss because of its comparable performance and its lighter training duty. The 2.9\% top-1 improvement from $(d)$ to $(c)$ indicates the effectiveness of our local-global graph Transformer encoder. Comparing $(c)$ and $(e)$, we could see that the local-global decoder achieves higher top-1 accuracy than the full global decoder, whereas the latter version has better top-$k$ accuracy for $k>1$. This is also reasonable. Focusing on a specific reaction center makes the generative process more constrained. The performance drop for $k>1$ indicates the loss of outcome diversity. However, the local-global decoder is compatible with the reaction center search, whereas the full global decoder is not. With the $search$ strategy, the model can boost the top-$k$ accuracy and improves the top-10 accuracy by 4.9\% from $(e)$ to $(f)$.

\begin{figure*}[t]
    \centering
    % trim={<left> <lower> <right> <upper>}
    \includegraphics[width=0.9\textwidth, trim={0.5cm 1cm 0.5cm 1cm},clip]{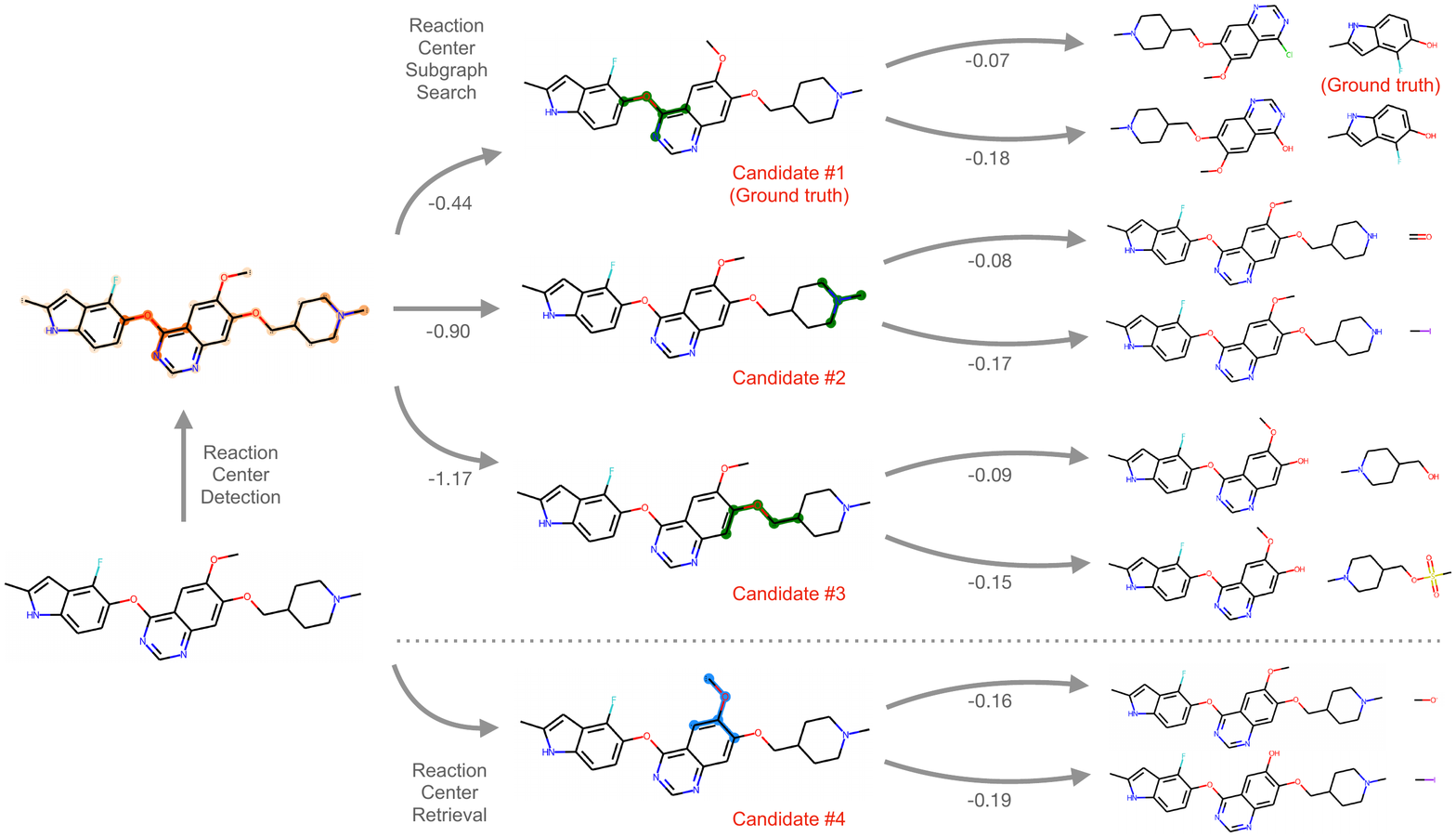}
    \caption{Generative procedure for a randomly selected molecule.}
    \label{fig:reaction_center}
\end{figure*}

\begin{figure}
    \centering
    % trim={<left> <lower> <right> <upper>}
    \subfigure[Success Case]{\label{fig:am_success}\includegraphics[width=\linewidth, trim={0.5cm 10.5cm 0.5cm 4.5cm},clip]{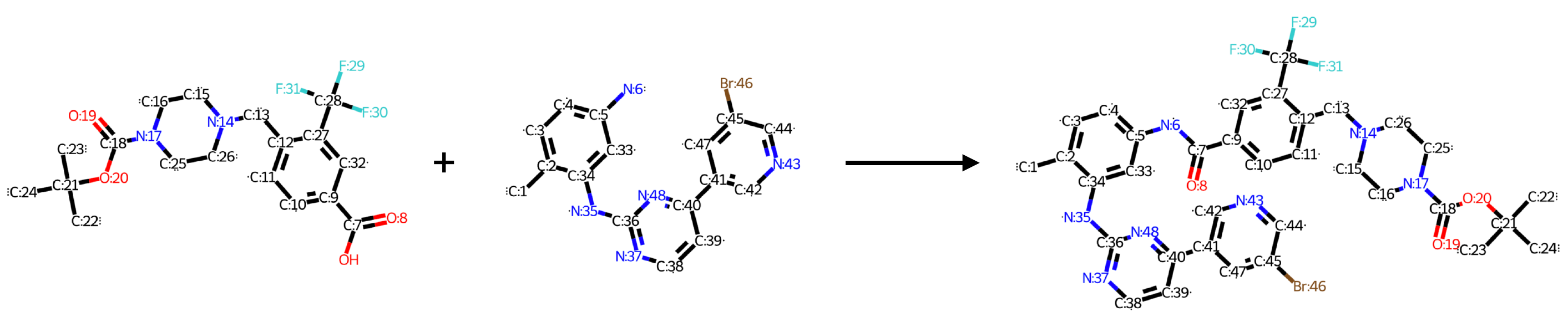}}
    \hfill
    \subfigure[Failure Case]{\label{fig:am_failure}\includegraphics[width=\linewidth, trim={0.5cm 12cm 0.5cm 5cm},clip]{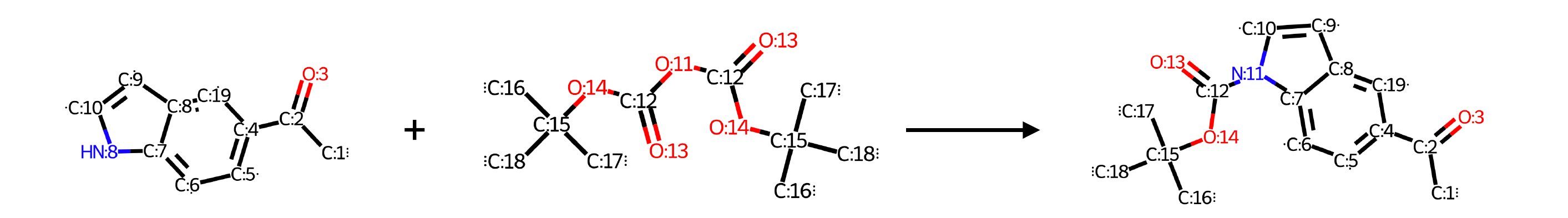}}
  \caption{Sample Atom Mapping.}
  \label{atom_mapping}
\end{figure}

\subsection{Qualitative Analysis}
\label{sec:qualitative}

In addition to its competitive benchmark performance, Retroformer is fully interpretable and controllable with external chemical guidance.

\paragraph{Reaction Center Detection}
To evaluate the interpretability and the quality of the detected reaction center, we randomly select a product molecule from the test set of USPTO-50K and predict the reactants with the $search$ strategy. We also evaluate the setting where we explicitly specify a reaction center and give it to the decoder. We term this setting as the reaction center retrieval. As shown in Figure~\ref{fig:reaction_center}, the $search$ algorithm proposes three different reaction centers (highlighted in green) given the raw reactive probabilities. The numbers represent the reactive scores of each subgraph candidate. In this example, the top candidate corresponds to the ground truth reaction center in the data. The third column indicates the top-2 verified predicted reactants given both the reaction center and the encoded molecule. The numbers represent the generative scores of each reactants. It shows that the model can understand the concept of reaction center and propose chemical transformations compatible with it. The outcomes from the retrieved reaction center (highlighted in blue) also demonstrate that the generation is fully controllable by external guidance. 
% interpretable, controllable, conditional
 
\paragraph{Atom Mapping}

Since our model is trained to learn the token alignment between the source and the target SMILES, the predicted attention can be easily converted to atom mapping. Different from the RXNMapper \cite{rxnmapper} that uses an additional neighbor attention multiplier to calculate the atom mapping from the attention weights, we directly use the attention weights to do the assignment while ignoring the molecular graphical structure. Note that it does not guarantee either the one-to-one mapping from product atom to reactants atom or the equivalence of the mapped element. Figure~\ref{fig:am_success} shows a success case of the inferred atom mapping. Figure~\ref{fig:am_failure} shows a typical failure case. This assignment mistakenly aligns [O:11] with [N:11]. The mistake is explainable since [HN:8] and [O:11] within the reactants are the exact position where chemical transformation happens. Also, this naive atom mapping fails to assign the one-to-one mapping, which is also reasonable because of the symmetry present in the second reactant. 

\section{Conclusion}
We propose Retroformer, a novel Transformer-based architecture that reaches the new state-of-the-art performance for template-free retrosynthesis. With the proposed local attention heads and the incorporation of the graph information, the model is able to identify local reactive regions and generate reactants conditionally on the detected reaction center. Being aware of the reaction center also encourages the model to generate reactants with improved molecule validity, reaction validity, and interpretability. We plan to further research the multi-step template-free retrosynthesis planning problem using Retroformer as the single-step retrosynthesis prediction backbone.

\medskip

\bibliography{references}
\bibliographystyle{acm}

\clearpage
\section{Appendix: SMILES Graph Construction}
\label{sec:smiles_graph}
To ensure the alignment between the SMILES token and the atoms in graph, we expand the original molecular graph $G_{mol}$ into the SMILES graph $G_{smi}$ by Algorithm~\ref{alg:smiles_graph}. $readSmiles()$, $getAtoms()$, $getNeighbors()$, and $writeSmiles()$ are functions supported in RDKit \cite{rdkit}. The tagging procedure is to inform the connection relationship between SMILES tokens. Rewriting the tagged SMILES without canonicalization is to ensure that the SMILES syntax does not change after the tagging. 

\begin{algorithm}[!htb]
   \caption{SMILES Graph Construction}
   \label{alg:smiles_graph}
    \begin{algorithmic}
       \STATE {\bfseries Input:} molecule canonical SMILES $S$
       \STATE Initialize $V$ as a token list of $S$.
       \STATE Initialize $E$ as an empty set.
       \STATE Initialize $A$ as an empty list.
       \STATE $M$ = $readSmiles$($S$)
       \FOR{$a_i \in getAtoms$($M$)}
            \STATE Assign tag \#1 to the SMILES symbol of $a_i$.
            \FOR{$a_j \in getNeighbors$($M, a_i$)}
                \STATE Assign tag \#2 to the SMILES symbol of $a_j$.
            \ENDFOR
            \STATE Get the tagged SMILES $S'$ = $writeSmiles$($M$, canonical=False)
            \STATE Retrieve the token connections $e_{(\#1, \#2)}$ by the tagged $S'$ and add them to $E$.
            \STATE Retrieve the bond feature of $e_{(\#1, \#2)}$ and add them to $A$.
       \ENDFOR
    \STATE {\bfseries Output:} $V$, $E$, $A$
    \end{algorithmic}
\end{algorithm}

\section{Appendix: SMILES Token Alignment Computation}
\label{sec:get_alignment}
The ground truth token alignment between the product SMILES and the reactants SMILES is computed as Algorithm~\ref{alg:get_alignment}. The algorithm takes the atom-mapped product and reactants as inputs. The computation works with both the canonical SMILES and the permuted SMILES.

\begin{algorithm}[!htb]
    \caption{SMILES Token Alignment Computation}
    \label{alg:get_alignment}
    \begin{algorithmic}
       \STATE {\bfseries Input:} atom-mapped product SMILES $S_p$ and atom-mapped reactants SMILES $S_r$.
       \STATE Initialize the token mapping dictionary $r2s$.
       \FOR{$s_{r_i} \in S_r$}
            \IF{$s_{r_i}$ is not visited and $s_{r}$ is an atom token}
                \STATE Locate the token $s_{p_j}$ in $S_p$ with the same atom mapping number as $s_{r_i}$: ${am}(s_{p_j}) == {am}(s_{r_i})$.
                \WHILE{$s_{r_i} == s_{p_j}$ or $am(s_{p_j}) == am(s_{r_i})$}
                    \STATE Add alignment relationship \{$i:j$\} into $r2s$.
                    \STATE Increment $i$ and $j$.
                \ENDWHILE
            \ENDIF
       \ENDFOR
       \FOR{\{$i:j$\} $\in r2s$}
            \STATE Decrement $i$ and $j$.
            \WHILE{$s_{r_i} == s_{p_j}$ and $s_{r_i}$ is not an atom symbol}
                \STATE Add alignment relationship \{$i:j$\} into $r2s$.
                \STATE Decrement $i$ and $j$.
            \ENDWHILE
       \ENDFOR
       \STATE {\bfseries Output:} $r2s$.
       
    \end{algorithmic}
\end{algorithm}

\begin{figure*}[!htb]
    \centering
    % trim={<left> <lower> <right> <upper>}
    \includegraphics[width=1.0\textwidth, trim={0.5cm 2cm 0.5cm 2cm},clip]{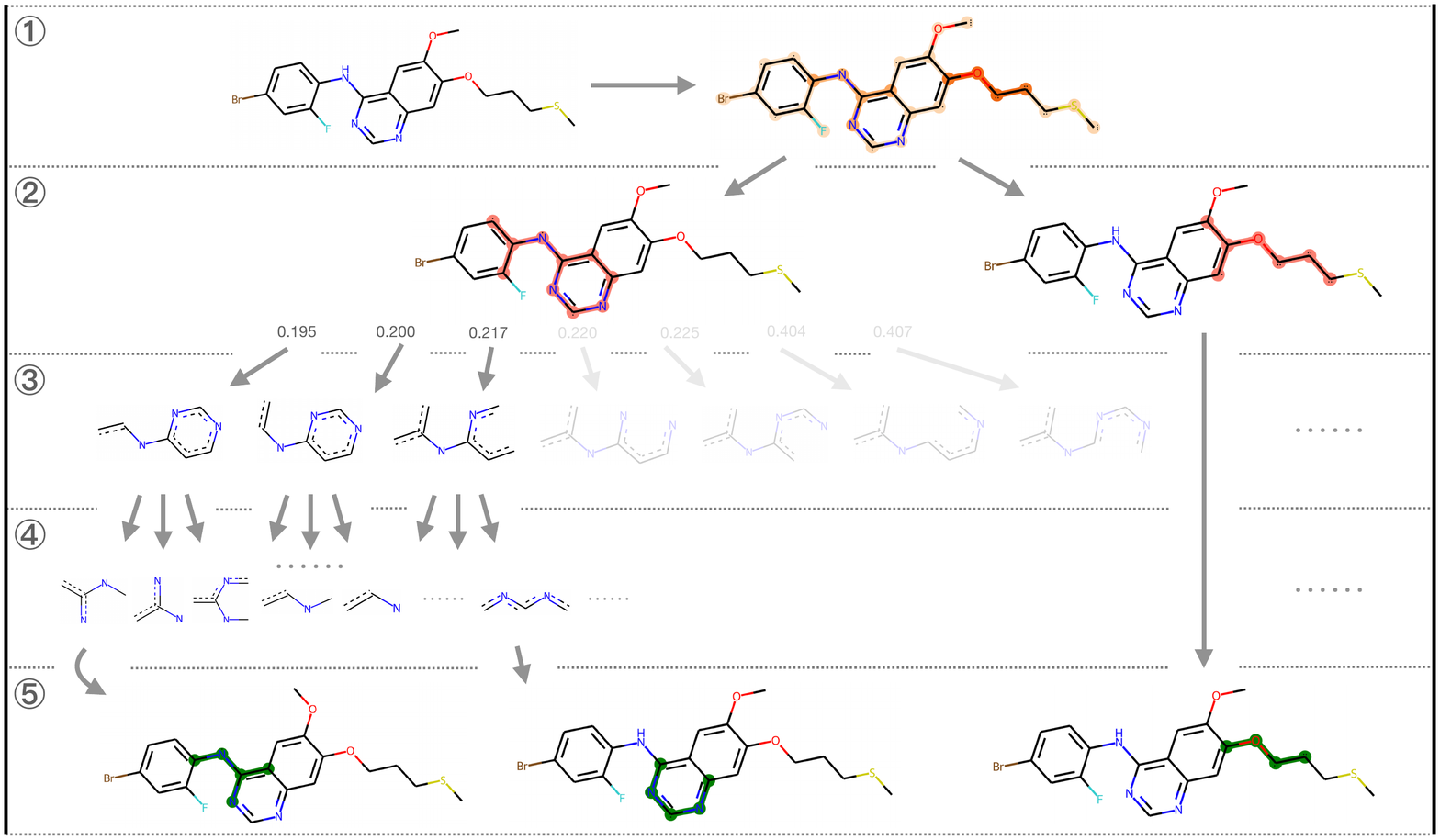}
    \caption{Visualization of the reaction center subgraph search algorithm: (1) Predict raw atom and bond reactive probability (i.e., reaction center detection); (2) Retrieve connected components based on the reactive probability; (3) Iterative pruning with $maxBranch=3$; (4) Retrieve and rank all the candidate subgraphs (i.e. reaction centers); (5) Select the top-$n$ diverse candidates from all subgraphs.}
    \label{fig:reaction_center_search}
\end{figure*}

\section{Appendix: Reaction Center Subgraph Search}
\label{sec:reaction_center_search}
Algorithm~\ref{alg:subgraph_search} shows the detailed $search$ reaction center subgraph search algorithm, and Figure~\ref{fig:reaction_center_search} shows a visualization of its procedure. In general, it searches for the candidate subgraphs (i.e., reaction centers) within the molecular graph via recursive pruning. It adopts a set of hyperparameters to avoid searching over the entire subgraph space.

The reaction center subgraph search is done in multiple stages: First, the algorithm removes all nodes whose $P_{rc}(s_i) < \alpha_{atom}$ and edges whose $P_{rc}(e_{ij}) < \alpha_{bond}$, and retrieves all the connected components $C$ from the edited graph. Second, for each connected component $c = (V_c, E_c)$, the algorithm retrieves all its subgraphs and the corresponding reactive scores via recursive pruning. The recursive pruning takes $maxRootSize, minLeafSize, maxBranch$ as three arguments to control its search space. If the number of nodes $|V_c|$ is larger than $maxRootSize$, then the algorithm directly removes $|V_c| - maxRootSize$ number of nodes with the lowest reactive scores. At each iteration, the algorithm considers nodes that lie along the border of the current graph as pruning candidates. $maxBranch$ is the maximum branching factor of the recursive pruning. The algorithm first ranks the pruning candidates by their atom reactive probability, and keeps only the top-$maxBranch$ candidates for pruning. The recursion stops when $|V_c| = minLeafSize$. After all subgraphs are retrieved from a root connected component $c$, we rank them by their reactive scores. Prior to the overall subgraph ranking, we remove all subgraphs (excluding the top-1 subgraph) that share at least two common nodes with the top-1 subgraph to ensure the candidates' diversity. At last, we gather the remaining subgraphs from all the root connected components $C$ and retrieve the top-$n$ reaction center candidates by their reactive scores.

In our experiments, we set $n=3$. Note that it only guarantees the maximum amount of reaction center candidates. We set the temperature $T=10$ to flatten the reactive probabilities $P_{rc}(s)$ and $P_{rc}(e)$. As for $\alpha_{atom}$ and $\alpha_{bond}$, instead of having a fixed value, we dynamically set the two parameters as the $k_s^{th}$ and $k_e^{th}$ percentile of $P_{rc}(s)$ and $P_{rc}(e)$, respectively. Based on the best validation performance, we set $k_s = 40, k_e = 40$ for reaction class unknown setting, and $k_s = 40, k_e = 55$ for reaction class known setting; $\beta$ is the parameter that controls the $minLeafSize$. For simplicity reason, we set $\beta = 0.5, maxRootSize = 25,$ and $maxBranch = 5$.

\begin{algorithm}[!htb]
    \caption{Reaction Center Subgraph Search}
    \label{alg:subgraph_search}
    \begin{algorithmic}
       \STATE {\bfseries Input:} $P_{rc}(s), P_{rc}(e), G_{smi}, \alpha_{atom}, \alpha_{bond}, \beta$.
       \STATE Remove all nodes whose $P_{rc}(s_i) < \alpha_{atom}$ from $G_{smi}$.
       \STATE Remove all edges whose $P_{rc}(e_{ij}) < \alpha_{bond}$ from $G_{smi}$.
       \STATE Retrieve all the connected components $C$ from the edited SMILES graph.
       \FOR{$c=(V_c, E_c) \in C$}
            \STATE Set $maxRootSize = 25$.
            \STATE Set $maxBranch = 5$.
            \STATE Set $minLeafSize = \sum_{s_i \in c} \mathbbm{1}(P_{rc}(s_i) > \beta)$, where $s_i \in c$. 
            \IF{$|V_c| > maxRootSize$}
                \STATE Remove $|V_c|-maxRootSize$ nodes with the lowest $P_{rc}(s_i)$.
            \ENDIF
            \STATE Retrieve all subgraphs of $c$ (with scores) via recursive pruning with $maxBranch$ and $minLeafSize$.
            \STATE Remove all subgraphs who share more than two common nodes with the top-1 subgraph.
       \ENDFOR
    \STATE {\bfseries Output:} Subgraphs with reactive scores.
    \end{algorithmic}
\end{algorithm}

\begin{figure}[!htb]
    \centering
    % trim={<left> <lower> <right> <upper>}
    \includegraphics[width=0.65\textwidth]{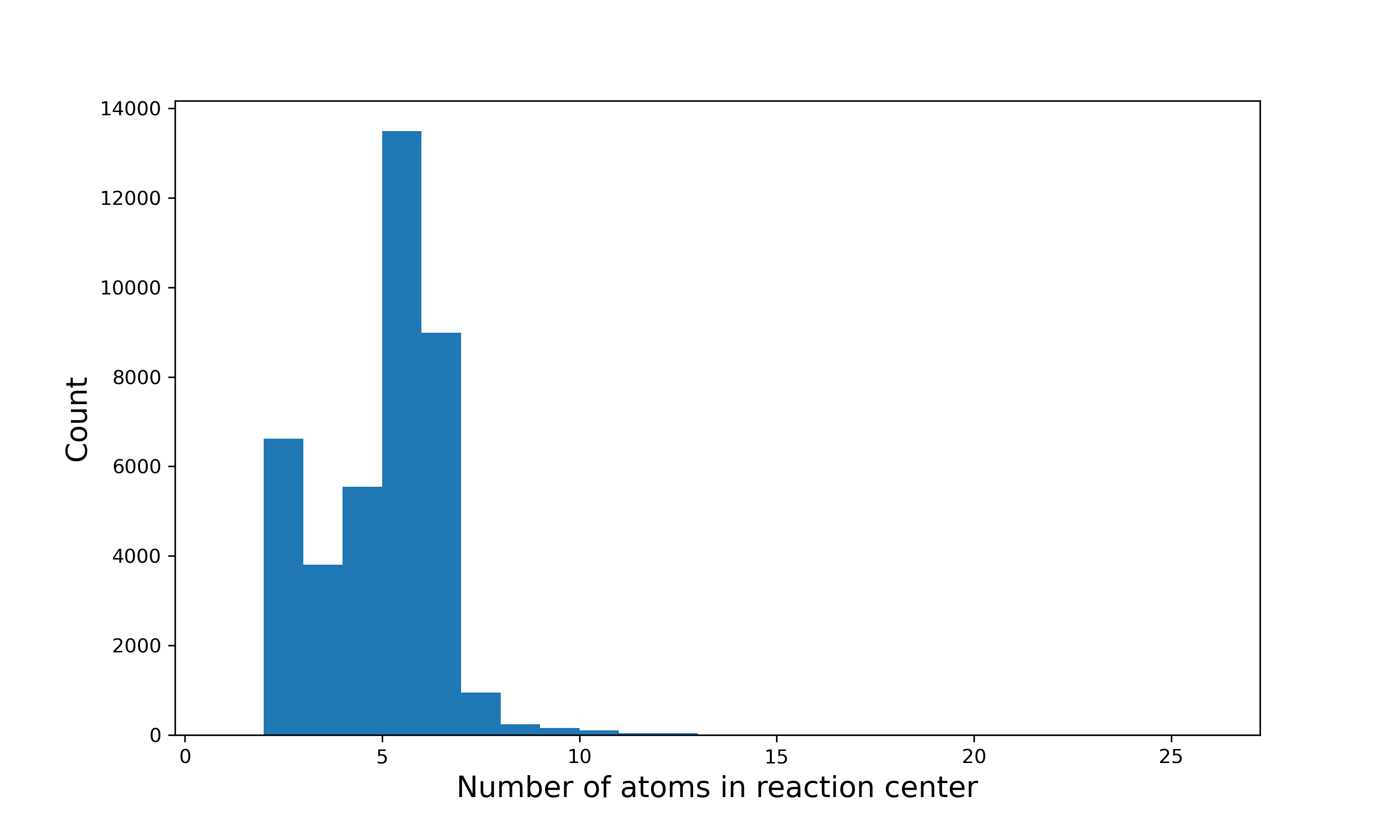}
    \caption{Histogram of the size of reaction centers in the training set of USPTO-50K.}
    \label{fig:reaction_center_hist}
\end{figure}

The exact reactive score for a candidate subgraph $G=(V, E)$ is computed as follow: 

\begin{align}
    \frac{1+\varphi(|V|, \mu, \sigma^2)}{M} \big(\sum_{s_i \in V} \log P_{rc}(s_i) + \sum_{e_{ij} \in E} \log P_{rc}(e_{ij})\big)
\end{align}

where $M = |V| + |E|$ is the normalization factor, and $\varphi(.)$ is the density function of a normal distribution of the size of reaction centers. We set $\mu = 5.55$ and $\sigma = 1.2$, which are computed from the training dataset. This is a heuristic factor taken from the observation that the size of the reaction centers has little relationship with the size of the molecule, but is rather normally distributed (Figure~\ref{fig:reaction_center_hist}). 

\clearpage
\section{Appendix: Bond Features}
\label{sec:bond_feature}

Table~\ref{tab:bond_feat} shows the bond features considered in the proposed Retroformer.
\begin{table}[!h]
    \centering
    \caption{Bond features.}
    \begin{tabular}{lll}
    \Xhline{1pt}
    \textbf{Bond Feature} & \textbf{Possible Values}         & \textbf{Size} \\ \hline
    Bond Type             & Single, aromatic, double, triple & 4             \\
    Aromatic              & True, false                      & 1             \\
    Conjugated            & True, false                      & 1             \\
    Part of Ring          & True, false                      & 1             \\ 
    \Xhline{1pt}
    \end{tabular}
    \label{tab:bond_feat}
\end{table}

\end{document}